\documentclass{article}
\usepackage{amsfonts}

\usepackage{amscd,amsmath,amssymb}

\newcommand{\p}[1]{(\ref{#1})}

\newcommand{\nn}{\nonumber}
\newcommand{\be}{\begin{equation}}
\newcommand{\ee}{\end{equation}}
\newcommand{\bea}{\begin{eqnarray}}
\newcommand{\eea}{\end{eqnarray}}
\newcommand{\ba}{\begin{array}}
\newcommand{\ea}{\end{array}}
\newcommand{\disty}{\displaystyle}

\begin{document}
\thispagestyle{empty}
\vspace*{1cm}
\begin{center}
{\Large\bf N=4, d=1 tensor multiplet and hyper-K\"ahler $\sigma$-models }
\end{center}
\vspace{1cm}

\begin{center}
{\large\bf S.~Krivonos, A.~Shcherbakov}
\end{center}

\begin{center}
{\it Bogoliubov Laboratory of Theoretical Physics, JINR, 141980 Dubna,
Russia}\\
\vspace{0.2cm}
{\tt krivonos, shcherb@theor.jinr.ru}
\end{center}
\vspace{1cm}

\begin{abstract}
\noindent We demonstrate how hyper-K\"ahler manifolds arise from a sigma-model action for
$N=4$, $d=1$ tensor supermultiplet after dualization of the auxiliary
bosonic component into a physical bosonic one.
\end{abstract}
\vspace{1cm}
\section{Introduction}
One of the interesting features of the $N=4$ supersymmetric mechanics theories is the diversity
of off-shell supermultiplets. It is shown in \cite{{GR},{PT}} that the minimal supermultiplets,
which contain four fermionic components, may have a different number of physical bosons: any natural
number from zero till four. Moreover, in one dimension one may switch between different supermultiplets
by expressing the auxiliary components through time-derivative of physical bosons, and vice versa.
This procedure is a pure algebraic one
and it may be used without any reference to actions. Basing on these peculiarities, in \cite{FG} the
term ``root supermultiplet'' was proposed for $N=4$, $d=1$ supermultiplet with four physical bosons and
four fermions. Such off-shell supermultiplets, which do not contain any auxiliary components exist only
in one dimension. Many of known supermultiplets with a smaller number of physical bosons can be obtained
{}from the ``root'' one by a proper reduction. The geometric nature of the relations between different
supermultiplets was clarified in \cite{ikl1} where a wide set of off-shell $N=4$ supermultiplets
was deduced by using nonlinear realizations of $N=4,d=1$ superconformal group. It turns out that
the physical bosonic components of these supermultiplets parameterize different coset spaces of the same
superconformal group. Finally, in \cite{bkmo} the relations between different supermultiplets were
extended to the level of actions and the term ``root'' action was invented just for the most general
action of the ``root'' supermultiplets. It was explicitly demonstrated that the general actions
for the supermultiplets with a smaller number of physical bosons might be obtained via a proper reduction procedure
{}from the ``root'' one.

The key observation, used in establishing the relations between different supermultiplets,
is very simple. Roughly speaking, if the transformation properties of some bosonic auxiliary
component $A$ of $N=4$ supermultiplet under supersymmetry read
\be\label{a}
\delta A \sim \mbox{ parameter} \times \partial_t ( \mbox{ physical fermions }),
\ee
then one may introduce a new physical bosonic component $u$ as $\partial_t u = A$
with the transformations law
$$
\delta u \sim \mbox{ parameter} \times ( \mbox{ physical fermions }).
$$
Then one may replace the auxiliary component $A$ by a new bosonic field $u$.
After such a replacement in the action, the term quadratic in the auxiliary component $A$ turns into
kinetic term for the field $u$. As a result, we get the supermultiplet with
one additional physical boson $u$. Clearly enough that if the field $u$ enters
the action only through its time derivative and, therefore, the action possesses isometry
with the Killing vector $\partial_u$, one may reverse the consideration above and
replace $\partial_t u$ by the new auxiliary component $A$.

Although being intuitively transparent, the discussed procedure has a loophole. It may happen
that for some supermultiplet there exists another expression $B$ which also starts with the auxiliary component
and transforms as a full time derivative under supersymmetry but nevertheless is essentially different from $A$.
This fact has no immediate consequences for the given supermultiplet\footnote{Observe, that one may use these
new expressions as the new additional Fayet-Iliopoulos terms in the action(see e.g. \cite{bkmo}).}, because
$B$ is algebraically related to the former auxiliary component $A$. However, if we decide to dualize $B$ instead of $A$ into a physical component, the resulting supermultiplet will be completely different.

In the present Letter we demonstrate that for the $N=4$, $d=1$ tensor supermultiplet with three physical bosons,
one auxiliary component, and four fermions there exists a functional freedom in choosing an
auxiliary component with the transformation properties \p{a}. After dualization of this new
auxiliary component into a physical boson we end up with the nonlinear off-shell $N=4$ supermultiplet with
four bosons and four fermions -- a new nonlinear variant of the ``root'' supermultiplet. The most interesting
result we achieved is that with some natural restrictions on the metric of the general sigma-model type action
for the tensor supermultiplet, the resulting bosonic action after dualization coincides with the one dimensional
variant of the general Gibbons-Hawking solution for four-dimensional hyper-K\"ahler metrics with one triholomorphic
isometry \cite{GH}.

Summarizing, the $N=4$, $d=1$ tensor supermultiplet may be dualized into new nonlinear supermultiplets with
four bosonic and four fermionic components. The proper constraints imposed on the metric of the general
sigma-model action for the tensor supermultiplet result in the hyper-K\"ahler type sigma model in the
bosonic sector of the dualized system.

\section{Preliminaries: $N=4$, $d=1$ tensor supermultiplet}
The simplest description of the $N=4$, $d=1$ tensor supermultiplet is achieved
by introducing a real triplet of $N=4$ superfields $V^{ab}$
\be\label{def1}
V^{ab}=V^{ba}, \quad \left(V^{ab}\right)^\dagger = V_{ab}, \quad a,b=1,2,
\ee
depending on the coordinates of the $N=4$, $d=1$ superspace $\Bbb{R}^{1|4}$
$$
\Bbb{R}^{1|4}=(t,\theta_{a}, \bar\theta{}^b), \quad \left( \theta_{a} \right)^\dagger = \bar\theta{}^a,
$$
where $a$, $b$ are doublet indices of $SU(2)$ group. The constraints which identify the multiplet are \cite{IS1}
\be\label{def2}
D^{(a}V^{bc)}=0, \quad {\bar D}{}^{(a}V^{bc)}=0,
\ee
where the covariant spinor derivatives $D^a, {\bar D}_a$ are defined by\footnote{We use the following convention for the
skew-symmetric tensor
$\epsilon$: $\epsilon_{ab}\epsilon^{bc}=\delta_a^c$, $\epsilon_{12}=\epsilon^{21}=1$. As usual, the round
brackets imply symmetrization of the enclosed indices with the normalizing factor 1/2.}
$$
D^a = \frac{\partial}{\partial \theta_a}+i \bar\theta{}^a \partial_t, \qquad
{\bar D}_a = \frac{\partial}{\partial \bar\theta{}^a}+i \theta_a \partial_t,\qquad
\left\{D^a,{\bar D}_b\right\}=2i\delta^a_b \partial_t.
$$
The constraints \p{def2} leave in $V^{ab}$ a real triplet of the dimensionless physical bosons $v_m$, four spinors $\lambda_a, \bar\lambda^b$
with dimensions \emph{cm}$^{-1/2}$, and the auxiliary field $A$ with dimension \emph{cm}$^{-1}$ that we define as
\be\label{compon}
v_m = \frac i2 \sigma^m_{ab} \; {V^{(ab)}}\vrule\,, \quad
\lambda_a = \frac 13 \bar D^b V_{ab} \vrule\,, \quad \bar\lambda^a = \frac 13 D_b V^{ab} \vrule\,, \quad
A = \frac i6 D^a \bar D^b V_{ab} \vrule\,,
\ee
where $\vrule$ means restriction to $\theta_a=\bar\theta{}^b=0$ and $\sigma^m_{ab}$ are the Pauli matrices. Under
$N=4$ supersymmetry these components transform as follows:
\be\label{tr1}
\delta v_m = i {\epsilon}^a \sigma^m_{ab} \bar\lambda^b - i \lambda^a \sigma^m_{ab} \bar{\epsilon}^b,\;
\delta \lambda_a = i {\epsilon}_a A + \sigma^m_{ab} \, {\epsilon}^b \, \dot v_m , \;
\delta \bar\lambda^a = - i \bar{\epsilon}^a A + \sigma^{m \; ab} \; \bar{\epsilon}_b \; \dot v_m ,\;
\delta A = {\epsilon}^a \dot{\bar\lambda}_a + \bar{\epsilon}^a \dot\lambda_a.
\ee
As regards the actions of the $d=1$ tensor multiplet, the general off-shell action
\be\label{action1}
S = \int {\mbox{d}} t {\mbox{d}}^2 \theta {\mbox{d}}^2 \bar\theta \, F(V^{ab}),
\ee
where $F(V^{ab})$ is an arbitrary real function of $V^{ab}$, is obviously invariant under
$N=4$ Poincar\'e supersymmetry. Being rewritten in terms of the components \p{compon} the action \p{action1}
reads
\be\label{action}
\ba{l}
\disty S = \int {\mbox{d}} t  \left[ g \left( \dot{v}_m \dot{v}_m + A^2
  + i ( \lambda^a \dot{\bar\lambda}_a - \dot\lambda^a {\bar\lambda}_a ) \right)
  + \partial_k g \left( - A \delta_{mk} - \varepsilon_{mnk} \dot v_n \strut \right)\lambda^a \sigma^m_{ab} {\bar\lambda}^b
  + \frac{\triangle g}4 \lambda^a \lambda_a {\bar\lambda}^b {\bar\lambda}_b \right].
\ea
\ee
Here, $\partial_m$ means differentiation with respect to $v_m$ and the metric $g$ is defined as
\be\label{g}
g \equiv \triangle F(v) = \frac{\partial^2 F(v)}{\partial v_m \partial v_m} .
\ee

The proper potential terms may be added to the action \p{action1} (see e.g. \cite{IL}). But what is important
for us here is that the bosonic sigma-model part of the action for the $N=4$ tensor supermultiplet forms a conformally
flat three-dimensional manifold with the metric given in eq.\p{g}.

\section{Dualization of the auxiliary component}
As we have already mentioned in the introduction, in one dimension one may switch between different
supermultiplets by a proper expression of auxiliary bosonic components
through time-derivatives of physical bosons, and vice versa. The reduction from the ``root'' supermultiplet
to the supermultiplets with a smaller number of the physical bosonic components was considered in detail
in \cite{bkmo}. Here we proceed in the opposite direction.
Our goal is to turn the auxiliary component $A$ which is present in the tensor supermultiplet $V^{ab}$ into a physical
boson and then to analyze the geometry of the resulting four-dimensional manifold.

The heart of the possible ``dualization" of the auxiliary component $A$ into physical ones lies in its transformation
properties under supersymmetry \p{tr1}. In one dimension one may integrate \p{tr1} and pass to the physical bosonic
component $u$ defined as
\be\label{u}
  \partial_t u = A
\ee
with the supersymmetry transformation properties
\be\label{tr3}
\delta u = {\epsilon}^a {\bar\lambda}_a + \bar{\epsilon}^a \lambda_a .
\ee
The resulting supermultiplet contains four physical bosonic and four fermionic components. The corresponding action
can be easily obtained from \p{action} using \p{u}
\be\label{naction}
\ba{l}
\disty S = \int {\mbox{d}} t  \left[ g \left( \dot{v}_m \dot{v}_m + \dot u^2
  + i ( \lambda^a \dot{\bar\lambda}_a - \dot\lambda^a {\bar\lambda}_a ) \right)
  + \partial_k g \left( - \dot u \delta_{mk} - \varepsilon_{mnk} \dot v_n \strut \right)\lambda^a \sigma^m_{ab} {\bar\lambda}^b
  + \frac{\triangle g}4 \lambda^a \lambda_a {\bar\lambda}^b {\bar\lambda}_b \right].
\ea
\ee
This action represents the particular type of the action for the so-called $N=4,d=1$ hypermultiplet \cite{{IL},{hm}}.
The restriction with respect to the general case
is that the metric $g$ depends on three bosonic fields $v_{m}$ and thus the action \p{naction} possesses
one obvious isometry with the Killing vector $\partial_u$.

One should stress that any auxiliary bosonic component which transforms as a full time derivative under supersymmetry
can be ``dualized'' into a physical field. So the question is whether the auxiliary component $A$ defined in \p{compon}
is unique? To answer this question let us consider the following easy-to-prove

{\tt Statement}
\emph{The most general real combination of dimension \emph{cm$^{-1}$} that is composed of the tensor supermultiplet
components, linear in its auxiliary component and transforms as a total time derivative has the following form
 (modulo total time derivative terms):
\be\label{combin}
B = a_m \dot v_m - f_{,m} \lambda^a \sigma^m_{ab} {\bar\lambda}^b + f A,
\ee
with
\be\label{combin1}
\delta B = {\epsilon}^a \frac{d}{dt} \left[ f {\bar\lambda}_a + i a_m \sigma^m_{ab} {\bar\lambda}^b \right]
  + \bar{\epsilon}^a \frac{d}{dt} \left[ f \lambda_a + i a_m \sigma^m_{ab} \lambda^b \right],
\ee
where real dimensionless coefficients $a_m$ and $f$ being functions of the vector $v_m$ satisfy
\be\label{restr}
\triangle f = 0, \qquad {\mbox{rot\,}} \vec a = \vec{\nabla} f.
\ee
}
Clearly enough, our previous choice of the auxiliary component $A$ corresponds to the very particular case
$f=1$. The vector $a_m$ is defined up to a gradient of a scalar function which is out of relevance here since this
corresponds to adding the total time derivative of the scalar function to $B$.

What is really important is that the newly defined auxiliary component $B$
is linear in the former auxiliary component $A$. This means that we can replace $A$ by the
proper combination which follows from (\ref{combin}) 
\be\label{repl}
A \rightarrow \frac{1}{f}\left[ B - a_m {\dot v}_m + f_{,m} \lambda^a \sigma^m_{ab} \bar\lambda^b \right].
\ee
The immediate consequences of this replacement are nonlinear transformations of the components.
Of course, while we are dealing with the $N=4$ tensor supermultiplet this nonlinearity is fake: we may
always come back to the auxiliary component $A$. However, this equivalence will be broken if we
 dualize the new auxiliary component $B$ as
\be\label{phi}
\partial_t \phi = B
\ee
with the following transformation properties under supersymmetry:
\be\label{ntr1}
\delta \phi = {\epsilon}^a \left[ f {\bar\lambda}_a + i a_m \sigma^m_{ab} {\bar\lambda}^b \right]
  + \bar{\epsilon}^a \left[ f \lambda_a + i a_m \sigma^m_{ab} \lambda^b \right]
\ee
The new supermultiplet with four bosonic components $v_m,\phi$ and four fermions
$\lambda^a, \bar\lambda{}^b$ is essentially nonlinear. The key point is that to pass to the previous
linear supermultiplet, one should solve the differential equation
\be\label{repl1}
\dot u = \frac{1}{f}\left[ \dot \phi - a_m {\dot v}_m + f_{,m}\lambda^a \sigma^m_{ab} \bar\lambda^b \right].
\ee
Thus, two equivalent formulations of the $N=4$ tensor supermultiplet being dualized give rize to two
different nonequivalent $N=4$ supermultiplets with four bosonic and four fermionic components: one is linear,
while the other supermultiplet is essentially nonlinear.

To clarify the role of the new physical bosonic field $\phi$, let us plug expression \p{repl}
with the auxiliary component $B$ dualized as in \p{phi} into action \p{action} and choose the metric $g = f$. The
bosonic part of the resulting action reads
\be\label{HK1}
S_{bosonic} = \int {\mbox{d}} t \left[ f \dot{v}_m \dot{v}_m + \frac 1f \left( \dot\phi - a_m {\dot v}_m\right)^2 \right].
\ee
Observe, this is just the one dimensional version of the general Gibbons-Hawking solution for four-dimensional
hyper-K\"ahler metrics with one triholomorphic isometry \cite{GH}. The corresponding Killing vector is obviously
$\partial_\phi$.

Thus, we explicitly demonstrate that starting from the sigma-model type action for the $N=4$ tensor supermultiplet one may recover
the hyper-K\"ahler sigma model by dualizing the proper auxiliary component. Let us remind that $N=4$ supersymmetric four dimensional
sigma model in one dimension does not oblige to have the hyper-K\"ahler sigma model in the bosonic sector. This fact is reflected
in the arbitrariness of the metric $g$. Only in the case $g=f$ with harmonic function $f$,
the bosonic part of the $N=4$ supersymmetric model becomes of hyper-K\"ahler type.

Summarizing, after dualization of the auxiliary component $B$ \p{combin} in the $N=4$, $d=1$ tensor supermultiplet we get a nonlinear
supermultiplet with four physical bosonic and four fermionic components which transform under $N=4$ supersymmetry
as follows:
\bea\label{fin1}
&& \delta v_m = i {\epsilon}^a \sigma^m_{ab} \bar\lambda^b - i \lambda^a \sigma^m_{ab} \bar{\epsilon}^b, \qquad
 \delta \phi = {\epsilon}^a \left( f {\bar\lambda}_a + i a_m \sigma^m_{ab} {\bar\lambda}^b \right)
  + \bar{\epsilon}^a \left( f \lambda_a + i a_m \sigma^m_{ab} \lambda^b \right), \nn \\
&& \delta \lambda_a = \frac if {\epsilon}_a \left( \dot\phi - a_m \dot v_m + f_{,m} \lambda^b \sigma^m_{bc} {\bar\lambda}^c \right) +
\sigma^m_{ab} \, {\epsilon}^b \, \dot v_m , \nn\\
&& \delta \bar\lambda^a = - \frac if \bar{\epsilon}^a \left( \dot\phi - a_m \dot v_m + f_{,m} \lambda^b \sigma^m_{bc} {\bar\lambda}^c \right) +
\sigma^{m \; ab} \; \bar{\epsilon}_b \; \dot v_m ,
\eea
with the functions $f(v^m)$ and $a_m(v^n)$ obeying to constraints
\be\label{fin2}
\triangle f = 0, \qquad {\mbox{rot\,}} \vec{a} = \vec{\nabla} f .
\ee
With the choice $g=f$ in the general sigma-model type action for tensor supermultiplet \p{action} it is dualized into
the action
\bea\label{fin3}
&& S =  \int {\mbox{d}} t \left[f \dot v_m \dot v_m + \frac 1f \left( \dot\phi - a_m {\dot v}_m\right)^2
  + i f \left( \lambda^a \dot{\bar\lambda}_a - \dot\lambda^a {\bar\lambda}_a \right) + \right. \nn \\
  && \qquad \left. + \frac 1f \left( \dot\phi - a_m {\dot v}_m\right) f_{,n} \lambda^a \sigma^n_{ab} {\bar\lambda}^b
  - {\varepsilon}_{mnk} \dot v_m f_{,n} \lambda^a \sigma^k_{ab} {\bar\lambda}^b
   \right],
\eea
which contains the hyper-K\"ahler sigma model in the bosonic sector.

\end{document}